\documentclass[reprint,superscriptaddress,showpacs,amsmath,amssymb,prb]{revtex4-1}

\usepackage{mathtools}
\usepackage{natbib}
\usepackage{siunitx}
\usepackage{array}
\usepackage{xcolor}
\usepackage{subfigure} 
\usepackage{amsmath} 
\usepackage{amsfonts}
\usepackage{amssymb}
\usepackage{graphicx}
\usepackage{hyperref}

\providecommand{\vect}[1]{{\boldsymbol{#1}}}

\providecommand{ }[1]{%
  {\textcolor{orange}{#1}}
  }

\begin{document}

\title{Nonlinear dynamics of topological ferromagnetic textures for frequency multiplication}

\author{D.R.~Rodrigues}
\affiliation{Institute of Physics, Johannes Gutenberg-Universit{\"a}t, 55128 Mainz, Germany}
\affiliation{University of Duisburg-Essen, Faculty of Physics, Lotharstrasse 1, 47057 Duisburg, Germany}
\author{J.~Nothhelfer} 
\affiliation{University of Duisburg-Essen, Faculty of Physics, Lotharstrasse 1, 47057 Duisburg, Germany}
\author{M.~Mohseni}
\affiliation{Fachbereich Physik and Landesforschungszentrum OPTIMAS, Technische Universit\"at Kaiserslautern,
67663 Kaiserslautern, Germany}
\author{R.~Knapman} 
\affiliation{Institute of Physics, Johannes Gutenberg-Universit{\"a}t, 55128 Mainz, Germany}
\author{P.~Pirro}
\affiliation{Fachbereich Physik and Landesforschungszentrum OPTIMAS, Technische Universit\"at Kaiserslautern,
67663 Kaiserslautern, Germany}
\author{K.~Everschor-Sitte}
\affiliation{University of Duisburg-Essen, Faculty of Physics, Lotharstrasse 1, 47057 Duisburg, Germany}
\date{\today}

\begin{abstract}
We propose that the non-linear radio-frequency dynamics and nanoscale size of topological magnetic structures associated to their well-defined internal modes advocate for their use as in-materio scalable frequency multipliers for spintronic systems. Frequency multipliers allow for frequency conversion between input and output frequencies, and thereby significantly increase the range of controllably accessible frequencies. In particular, we explore the excitation of eigenmodes of topological magnetic textures by fractions of the corresponding eigenfrequencies. We show via micromagnetic simulations that low-frequency perturbations to the system can efficiently excite bound modes with a higher amplitude. For example, we excited the eigenmodes of isolated ferromagnetic skyrmions by applying half, a third and a quarter of the corresponding eigenfrequency. 
We predict that the frequency multiplication via magnetic structures is a general phenomenon which is independent of the particular properties of the magnetic texture, and works also for magnetic vortices, droplets and other topological textures.

\end{abstract}

\maketitle

\section{Introduction} 
\label{Introduction}

\begin{figure}
\begin{center}
    \includegraphics[width=\columnwidth]{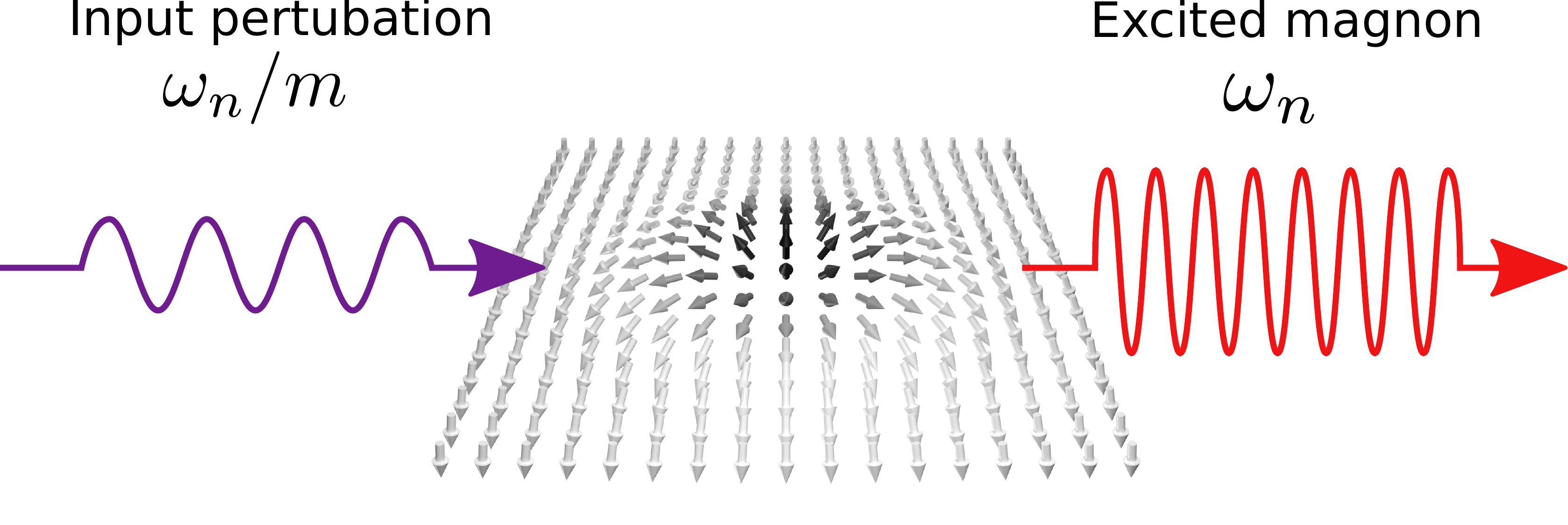}
    \caption{Sketch of an example of frequency multiplication using a magnetic skyrmion. Perturbations with a fraction of the eigenfrequency $\omega_{n}$ can excite the corresponding eigenmode of the skyrmion. Moreover, when the input perturbation are created by an AC magnetic field, the amplitude of the exited eigenmode can be larger than the amplitude of the input magnon.}
    \label{fig:FrequencyMultiplier}
    \end{center}
\end{figure}

Frequency multiplication is an important phenomenon of nonlinear oscillators from which one obtains high frequency outputs given low frequency inputs. It has a broad use in communication circuits~\cite{quievy1990,sanderford1998,kool1999,salvi1999,chien2000,chattopadhyay2004}, optical experiments~\cite{Miller1964,Guyot-Sionnest1986,Shen1989,Campagnola2011,Chen2012} as well as spintronic devices~\cite{Fiebig2002,Bibes2007,Bykov2012,Sebastian2013,Nemec2018,Leroux2020}. In spintronic experiments, different methods going beyond pure spintronics techniques are often employed to controllably create and manipulate magnons at different frequencies.
We predict that the nonlinear dynamics of topological magnetic structures provides an in-materio scalable pure spintronics-based alternative for frequency multipliers, see Fig.~\ref{fig:FrequencyMultiplier}.

The excitation modes of magnetic vortices~\cite{Shinjo2000,VanWaeyenberge2006,Dussaux2010}, skyrmions~\cite{Nagaosa2013,Buttner2015,Fert2017,Lonsky2020}, and droplets~\cite{Mohseni2013,Macia2014,Iacocca2014,Sulymenko2018} have relevant applications in magnetic tunnel junctions~\cite{Dussaux2010,Nishimura2002},  racetrack memories~\cite{Tomasello2014,Carpentieri2015}, microwave generators~\cite{Dussaux2010,Ruotolo2009}, and non-conventional computing~\cite{Torrejon2017,Bourianoff2018,Finocchio2019,Pinna2018}.
Excitation modes of ferromagnetic topological textures are bound magnon modes.~\cite{Vogt2011,Macia2014,Garst2017,Rodrigues2017a,Prokopenko2019,Mohseni2020,Lonsky2020}. They are typically classified by an integer number $n$ associated to a quantized angular momentum: i) breathing modes ($n=0$)~\cite{Mohseni2013,Kim2014d,McKeever2018,Penthorn2019}, ii) gyration modes ($|n|=1$)~\cite{VanWaeyenberge2006,Schwarze2015,Schutte2014}, and iii) higher order modes ($|n|>1$)~\cite{Vogt2011,Schutte2014,Rodrigues2017a,Kravchuk2017a}, see Fig.~\ref{fig:ExcitationModes}. 
The sign of $n$ determines the rotation direction, clockwise or counterclockwise. So far, most applications of skyrmions rely on linearly approximating the dynamics.  These excitation modes, however, are fundamentally non-linear. This implies, for example, an amplitude dependence of the excitation mode frequencies, as well as the existence of harmonic generation, i.e. the excitation of integer multiples of applied frequencies.

In general, the nonlinearity of magnetization dynamics is associated with many-magnon scattering~\cite{Fleury1968,Hurben1998,Zakeri2007,Schultheiss2009,Schultheiss2012,Chumak2014,Okano2019}. An example of the use of nonlinear phenomena creating multiple magnons is parallel pumping~\cite{Melkov2001,Sandweg2011,Ando2011c,Edwards2012,Bauer2015,Bracher2017a,Verba2018} which has been demonstrated both theoretically and experimentally. Parallel pumping has been used to generate magnon Bose-Einstein condensates, for example~\cite{Giamarchi2008,Serga2014,Bracher2017a,Okano2019}. In this nonlinear process, associated with parametric excitation, one can excite a certain magnon mode by applying an AC field with twice the corresponding eigenfrequency.~\cite{Zakharov1975,Bryant1988,Bertotti2001} The high frequency of the required driving field as well as the resonant magnon excitation at the driving field's frequency are caveats of the experimental implementation of parallel pumping. Moreover, the resonance at the frequency of the driving field is very sensitive to defects~\cite{Melkov2001,Hals2010}. Interactions of magnons with topological magnetic textures provide new possibilities for controlled and robust many magnon scattering manipulation~\cite{Sheka2001,Schutte2014,Iwasaki2014,Finocchio2015,Garst2017,Lonsky2020,Tatara2020,Seki2020,Wang2020}.

In this work, we propose the excitation of topological magnetic textures via frequency multiplication. Applying a perturbation with a fraction of an eigenfrequency of a bound mode, leads to a resonance at the corresponding eigenfrequency, see Fig.~\ref{fig:FrequencyMultiplier}. In particular, we obtain the excitation of the breathing $n=0$ mode of isolated skyrmions by applying AC magnetic fields, both in-plane and out-of-plane, with half and a third of the corresponding eigenfrequency. We also analyze the amplitude dependence of the excited eigenmode in terms of the amplitude of the applied AC field and the damping parameter.
The resonance of the bound eigenmodes with perturbations at fractions of the eigenfrequencies presents several advantages for applications:
i) above a certain threshold amplitude for the applied field, the excitation with a fraction of the eigenfrequency is more efficient than with same frequency, i.e.\ the amplitude of the eigenmode is bigger when applying a fraction of the eigenfrequency than applying the eigenfrequency itself; ii) the perturbations with fractional frequency are not eigenstates of the system, and thus, decay quickly away from the topological texture not contributing significantly to instabilities; iii) the eigenfrequencies can be tuned by changing the material parameters, for example by temperature changes, or by applying constant magnetic fields. This means that for given a frequency, it is possible to design a topological texture that allows for a resonant frequency multiplication.

\begin{figure}
\begin{center}
    \includegraphics[width=\columnwidth]{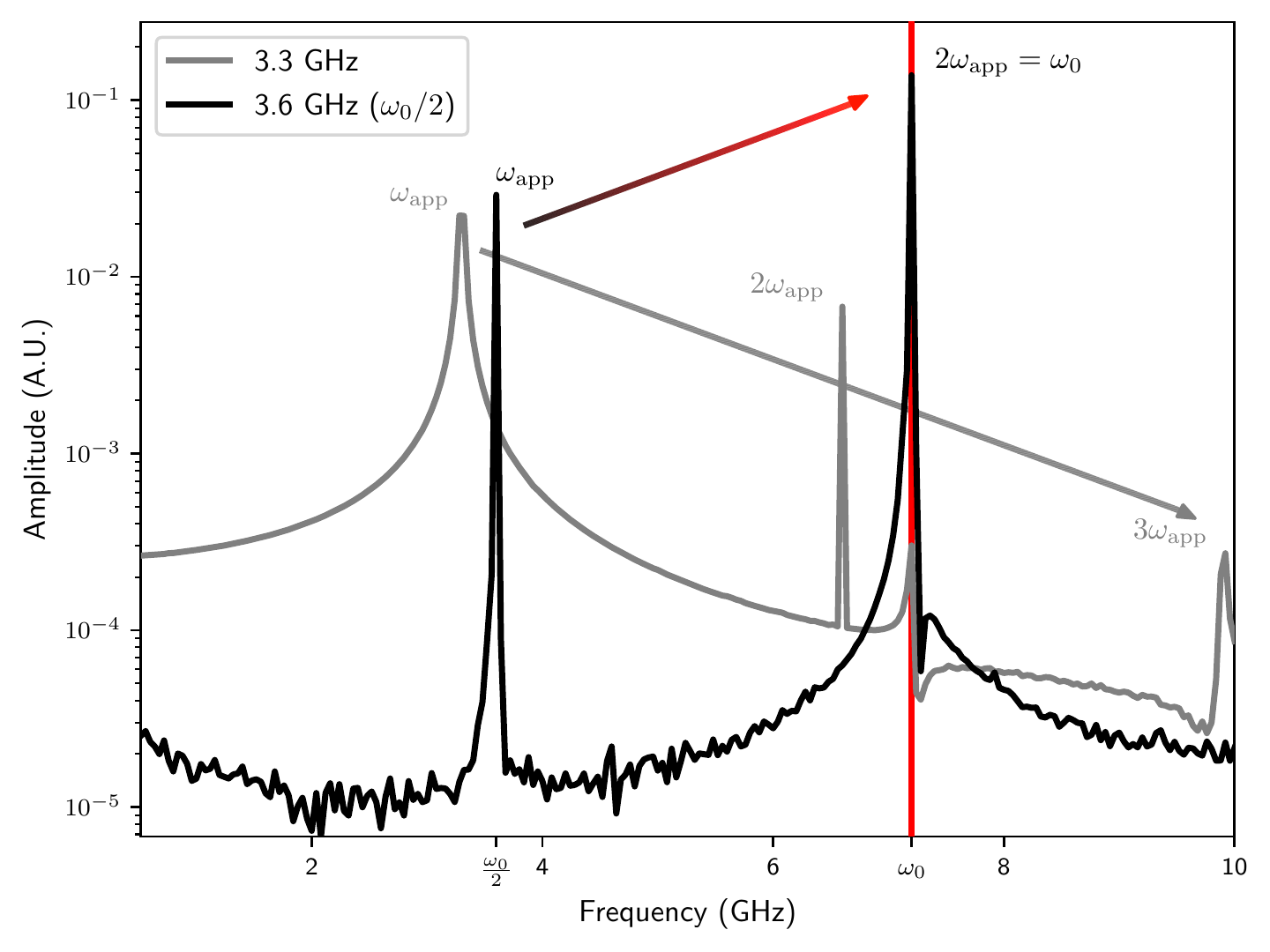}
    \caption{Illustrative example of general principles of excitation by frequency multiplication. We show the excitation of a skyrmion by magnetic fields with amplitude $B=5\,\mathrm{mT}$ at two different driving frequencies $\omega_{\textrm{app}}$. Due to the localized nonlinear potential of the skyrmion configuration, both driven excitations generate harmonics corresponding to integer multiples of the frequency. In the case in which the harmonic coincide with an eigenstate, there is resonance and the corresponding multiple has a higher amplitude.}
    \label{fig:IntuitivePic}
    \end{center}
\end{figure}

 {This manuscript is organized as follows. In Sec.~\ref{sec:Model} we discuss the general model, which we use to predict the possibility to excite eigenmodes of magnetic topological objects by fractions of its eigenfrequency. We describe the general case of frequency multiplication and the underlying physical behavior and assumptions. The theory is valid for topological objects in general and is expected to happen independently of the particular local topological magnetic configuration. In Sec.~\ref{sec:Skyrmion} we present frequency multiplication for excitation of skyrmions as an example. We show that it is possible to excite the skyrmion modes, including higher order modes that have not yet been experimentally studied. We show that the method of excitation by a fraction of the eigenmode can be more efficient than the usual linear excitation. Moreover, we study via micromagnetic simulations the efficiency of the frequency multiplication dependence on the amplitude of the driving force and the damping. We conclude in Sec.~\ref{sec:Discussion} by emphasizing the advantages of the frequency multiplication and discussing its difference compared to parametric excitation. Furthermore, we finalize by proposing applications for the method described in this manuscript.}

\section{General Model}
\label{sec:Model}

 {Frequency multiplication is a well known phenomenon in non-linear optics where it is also named higher harmonic generation~\cite{Franken1961,Lupke1999,Downer2001}.} It relies on the non-linear properties of the systems and can be explained in a general picture. Considering a field $\phi$ of a nonlinear system, the dynamics of small perturbations $\tilde{\phi}(t) \ll 1$ of the static background $\phi_{0}$ can be expressed as the expansion
\begin{equation}\label{eq:nonlinear}
\frac{d\tilde{\phi}}{dt} \approx \mathcal{L}_{1}(\phi_{0})\tilde{\phi}+ \mathcal{L}_{2\,}(\phi_{0})\tilde{\phi}^2 + \dots + \mathcal{L}_{p}(\phi_{0})\tilde{\phi}^{p} + \cdots\,.
\end{equation}
The first term on the right $\mathcal{L}_{1}\tilde{\phi}$ corresponds to a linear approximation for the nonlinear system and provides the eigenstates of the system with frequencies $\omega_{n}$. Terms $\mathcal{L}_k$ with $k>1$ correspond to interactions between the perturbations. They renormalize the value of the eigenfrequencies $\omega_{n}$ and lead to amplitude-dependent frequencies. If we consider a perturbation with a fraction of an eigenfrequency, such as $ \tilde{\phi}(t) \approx \phi_{\omega_{n}/2}\cos(\omega_{n}t/2)$, the quadratic term generates a contribution $\tilde{\phi}(t)^2\approx \tilde{\phi}_{\omega_{n}/2}^{2}\cos(\omega_{n}t)$ which corresponds to a solution to the linear term. This principle can be extended to $m>2$ and reveals why fractions of the eigenfrequency may lead to the excitation of the corresponding eigenmode.  {In general, the amplitude of the eigenmode $\omega_{n}$ grows with the $m$th-power of the driving amplitude with frequency $\omega_{n}/m$. Moreover, the expansion \eqref{eq:nonlinear} is independent of the amplitude of the perturbation $\varphi$.}

In particular, the dynamics of the unitary magnetization $\vect{m} = \vect{M}/M_{s}$, where $M_s$ is the saturation magnetization, is well described by the Landau-Lifshitz-Gilbert (LLG) equation~\cite{Gilbert2004}
\begin{equation}
\frac{d\vect{m}}{dt} = - \frac{\gamma}{M_{s}}\, \vect{m} \times \vect{B}_{\mathrm{eff}} + \alpha\, \vect{m} \times \dot{\vect{m}}.
\label{eq:model:LLG}
\end{equation}
Here $\gamma$ is the gyromagnetic ratio, and $\vect{B}_{\mathrm{eff}}=-\delta E[\vect{m}] / \delta\vect{m}$ is the effective magnetic field in a system with total magnetic energy density $E[\vect{m}]$.  {The non-linearity of Eq.~\eqref{eq:model:LLG} allows for the existence of solitons corresponding to localized stable non-collinear magnetic configurations characterized by topological properties. Moreover, low frequency perturbations to these topological configurations are bound to the vicinity of the non-collinear structure and behave as localized traveling waves.~\cite{Vogt2011,Macia2014,Garst2017,Rodrigues2017a,Mohseni2020,Lonsky2020} The non-linear potential associated with these topological objects gives rise to frequency multiplication of driven perturbations. If we consider perturbations $\delta\vect{m}$ of the topological configuration, given by $\vect{m}_{0}$, we can expand the LLG Eq.~\eqref{eq:model:LLG} as Eq.~\eqref{eq:nonlinear} including higher orders of $\delta\vect{m}$.}

 {The exact expansion depends on the configuration $\vect{m}_{0}$ and its symmetries. From an intuitive picture, however, one can derive some general principles taking into account the traveling perturbations in the non-collinear region of the topological object, see Fig.~\ref{fig:IntuitivePic}. First, we consider that any low frequency perturbation $\delta \vect{m}$ around the non-collinear region of the configuration $\vect{m}_{0}$ behaves as an underdamped traveling wave given by}
\begin{align}
\delta \vect{m} = \sum_{n}\Big(\phi_{1}(r, \psi, t)\left(\vect{m}_{0}\times\hat{\vect{n}}\right)\notag\\
+ \phi_{2}(r, \psi, t)\left((\vect{m}_{0}\times\hat{\vect{n}})\times\vect{m}_{0}\right)\Big),
\end{align}
 {where $\phi_{1},\phi_{2} \ll 1$ are dynamically conjugated and can be described in terms of collective coordinates.~\cite{Malozemoff1979,Clarke2008,Rodrigues2017} Here, $r,\psi$ are polar coordinates, and $\hat{\vect{n}}$ corresponds to the direction of the field-polarized background. If the polar and time dependence of $\phi_{1\,n},\phi_{2\,n}$ are given by the combination $\omega_{n}(t - t_{0}) - n\psi$, with $n$ an integer, the perturbation corresponds to the eigenmode of the topological excitation with eigenfrequency $\omega_{n}$. Perturbations propagate in the localized nonlinear potential of the topological structure as an underdamped traveling wave. The frequency of the amplitude depends on the amplitude of the perturbation and spatial landscape of the potential. Since the damping has a viscous nature, faster perturbations experience a higher damping.}

 {Traveling waves propagating in nonlinear potentials naturally generate harmonics corresponding to excitations with integer multiples of the original frequency.~\cite{Franken1961,Lupke1999,Downer2001} When an harmonic of the driven perturbation coincide with the natural eigenfrequencies of the system, there is resonance and the corresponding eigenmode is excited, see Fig.~\ref{fig:IntuitivePic}. Since excitations with lower frequency are less damped, one expects that at a certain regime of amplitude of the driving field and range of damping values, the excitation by frequency multiplication to be more efficient than the linear resonance of the eigenmode.}

 {The basic assumption for the frequency multiplication is the existence of a localized non-linear potential associated to the non-collinear topological structure. The frequency multiplication happens for any driven perturbation. For this reason, there is no threshold for the driving field. Increasing amplitude of the driving perturbation however leads to a higher efficiency of non-linear process. In addition, selection rules may be derived from the spatial distribution of the driven perturbation. We consider a Fourier expansion of the perturbation in terms of the eigenmodes of the topological object, such as}
\begin{equation}
\phi_{i\,n}(r, \psi, t) = \sum_{n}\phi_{i\,n}(r)\phi_{n}(r, n\psi - \omega_{n}t),
\end{equation}
 {where $\phi_{n}$ are the eigenmodes. We claim that a perturbation can excite via frequency multiplication any mode with non-vanishing $\phi_{i\,n}(r)$, see Fig.~\ref{fig:ExcitationModes} for example. Note that, increasing the driving force may lead to an increase of amplitudes $\phi_{i\,n}(r)$ which are almost vanishing for smaller driving forces.}

 {A quantitative analysis depends on the specific magnetization dynamics and the underlying topological object. In the next section we consider the skyrmion as an example to demonstrate frequency multiplication in magnetic topological objects.}

\section{Example: magnetic skyrmions}
\label{sec:Skyrmion}

While the results above are quite general, in this manuscript we focus on the excitation modes of isolated skyrmions stabilized by perpendicular anisotropy~\cite{Rohart2013,Kravchuk2017a} without loss of generality. 
The excitation modes of skyrmions have been studied by analytical and numerical methods which considered both linearized equations of motion as well as full micromagnetic simulations~\cite{Lin2014,Iwasaki2014,Schutte2014,Garst2017,Kravchuk2017a,Rodrigues2017a}. The experimental excitation and detection of these internal modes, in particular modes with $|n| > 1$ are still a challenge~\cite{Garst2017,Pollath2019,Penthorn2019,Finco2020,Lonsky2020}. This is because some modes only couple at linear level with fields that obey certain spatial distributions and symmetries~\cite{Garst2017,Beg2017}. We demonstrate that bound modes can be excited by homogeneous alternating fields with fractions of the corresponding eigenfrequencies. Since the fractional excitation is more efficient for certain field amplitudes, this provides a path for the experimental excitation and detection of all skyrmion modes.

To obtain explicit results for an anisotropy stabilized isolated skyrmion we considered the following model 
\begin{equation}
\begin{split}
        E[\vect{m}] = \int d V  \,\,
     \big[ A \,(\nabla \vect{m})^2 - D \,\vect{m} \cdot \left(\left(\hat{\vect{z}}\!\times\!\nabla\right)\!\times\!\vect{m}\right) \\
     - K \,(m_z^2-1) \big],
\end{split}
\label{eq:model:energy}
\end{equation}
where $A$ is the magnetic stiffness, $D$ characterizes the strength of the interfacial Dzyaloshinskii-Moriya (DM) interaction, and $K$ is the strength of the effective uniaxial anisotropy which incorporates a correction due to a local approximation of the magnetostatic interactions~\cite{Bogdanov1994}. We consider $D < D_{c} = 4\sqrt{AK}/\pi$ such that the ferromagnetic state is the ground state of the system~\cite{Rohart2013,Kravchuk2017a,McKeever2018}. The energy contribution due to an external applied magnetic field $B_{\textrm{ext}}$  {is given by the Zeeman term}
\begin{equation}\label{eq:Zeeman}
E_{B_{\textrm{ext}}}[\vect{m}] = \mu_{0} B_{\textrm{ext}}(t) \int d V  \,\, \hat{\vect{n}}\cdot\vect{m}(\vect{x},t).
\end{equation}
 {where $\mu_{0}$ is the vacuum permeability.}

\begin{figure}[tb]
	\includegraphics[width=\columnwidth]{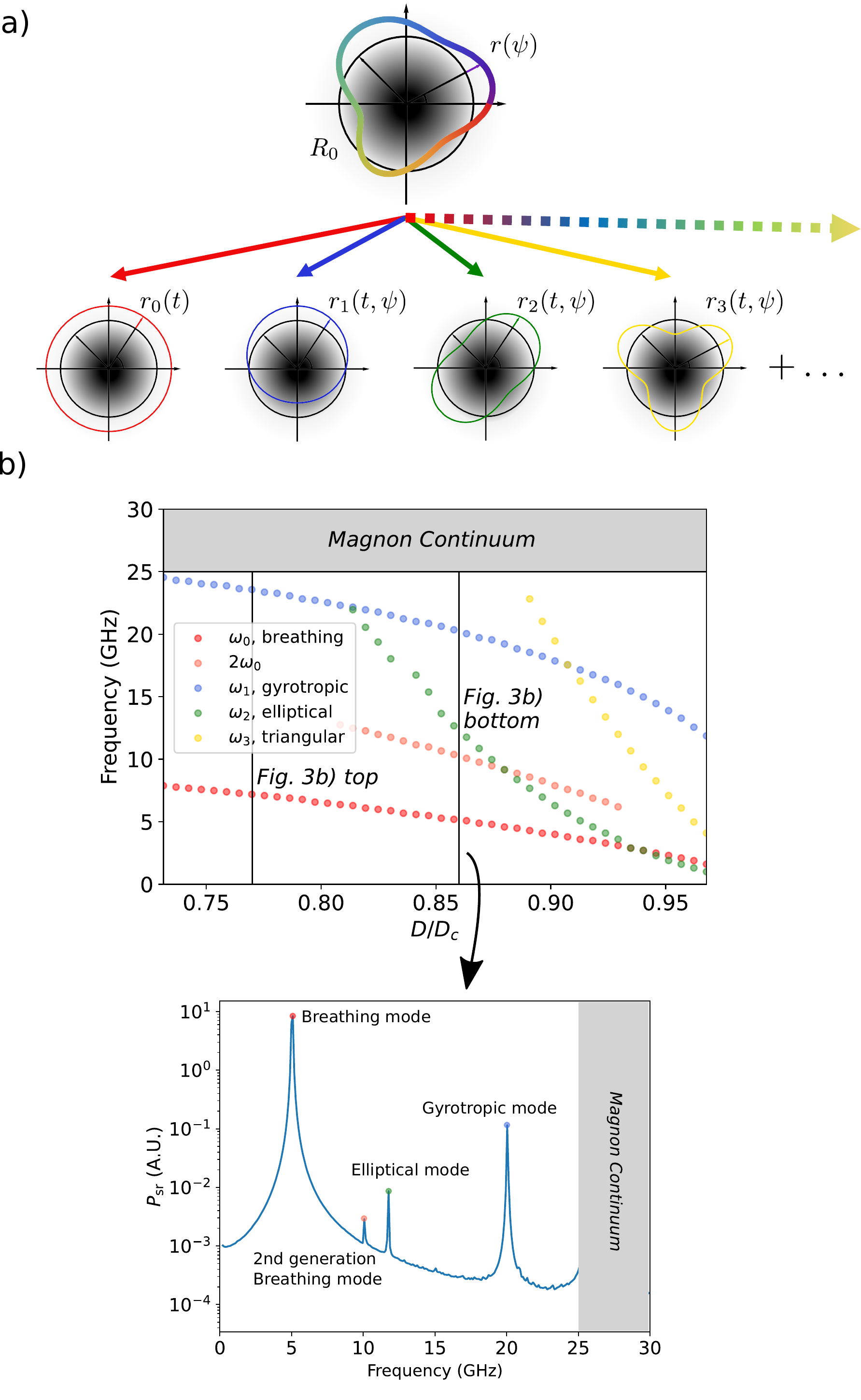}
	\caption{Excitation modes of skyrmions. a) Sketch of the decomposition of a smooth deformation into the internal skyrmion modes. The magnetization inside and outside the boundary have opposite directions, going from black to white. 
	$R_{0}$ is the ground state radius of the skyrmion, and, $r_{n}$ characterizes the bound modes. 
	The mode $n = 0$ represents the breathing mode. Modes with $|n|>0$ rotate (counter-)~clockwise with an amplitude-dependent frequency.
	b) Frequency spectrum of a Néel skyrmion as a function of rescaled DMI strength $D/D_c$ (top panel) and detailed analysis for a fixed DMI strength $D/D_c = 0.86$ (lower panel). The power spectrum in the lower panel was obtained by performing a Fourier transform of the magnetization dynamics~\cite{Kumar2017}.
	}  
	\label{fig:ExcitationModes}
\end{figure}

For a skyrmion modeled by Eq.~\eqref{eq:model:energy}, the spin wave eigenstates are given by the bound modes at the skyrmion with frequencies $\omega_{n}$ as well as a continuous distribution of magnon modes above the anisotropy gap with frequencies $\omega = (2\gamma/M_{s})(K + A k^2)$ where $k$ is the wave number~\cite{Kravchuk2017a}.
In Fig.~\ref{fig:ExcitationModes}b) we show the magnon spectrum obtained from micromagnetic simulations. Beyond the eigenmodes obtained from linearization calculations~\cite{Kravchuk2017a} we are able to identify the second harmonic of the breathing mode and a gyrotropic mode. It is important to notice that, for anisotropy stabilized skyrmions (at zero magnetic field), most modes exist below the magnon gap only for $D$ close to $D_{c}$~\cite{Kravchuk2017a,Rodrigues2017a}.
The micromagnetic simulations of this manuscript were performed with MuMax$3$~\cite{Vansteenkiste2014} using the following parameters:
$M_s= 1.1 \cdot 10^6 A/m$,  $A= 1.6 \cdot 10^{-11} J/m$, $K=5.1 \cdot 10^5 J/{m^3}$.

\begin{figure}
\begin{center}
    \includegraphics[width=0.7\columnwidth]{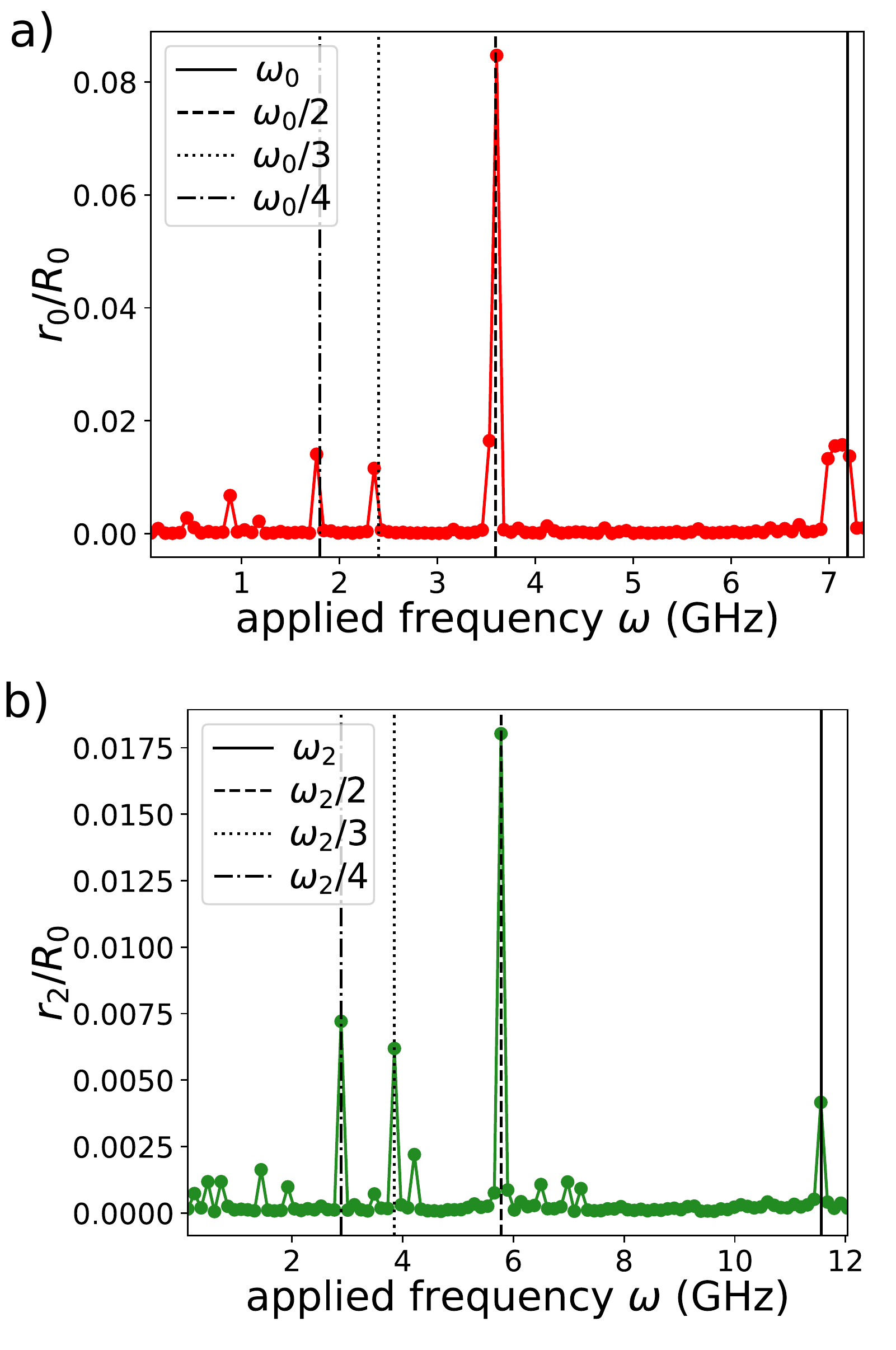}
    \caption{ Amplitude of the a) breathing ($\omega_0$), b) elliptical ($\omega_2$) mode as a function of the frequency of the in-plane applied field $B_{\mathrm{ext}}=0.05\, \mathrm{T}$ with $\alpha =10^{-3}$. Fractions of the corresponding frequency can still strongly excite the desired mode.  }
    \label{fig:Fractions}
    \end{center}
\end{figure}

To reveal the resonance with a fraction of the eigenfrequency, we computed the amplitude of the eigenmodes with frequency $\omega_{n}$ as a function of the applied frequency $\omega$ for an in-plane magnetic field excitation.
 {Perturbations of an isolated circular skyrmion driven by out-of-plane magnetic fields are radially symmetric and only couple to the breathing mode. In the case of an out-of-plane field applied to a skyrmion with no radial symmetry or an in-plane driving field generate perturbations that can excite any eigenmode.}
In Fig.~\ref{fig:Fractions} we show the results of exciting a) the breathing $n=0$ and b) the elliptical $n=2$ modes at different applied frequencies but same amplitude of the in-plane applied field.
We noticed that the eigenmodes are excited by applying fractions of the corresponding frequency, i.e.\ $\omega = \omega_{n}/m$ where $m \in \mathbb{Z}$. 
The same behavior was obtained for the triangular $n=3$ mode. The observation of resonance peaks at integer fractions of the corresponding eigenmodes is an important result of this manuscript. 

\begin{figure}
\begin{center}
    \includegraphics[width=\columnwidth]{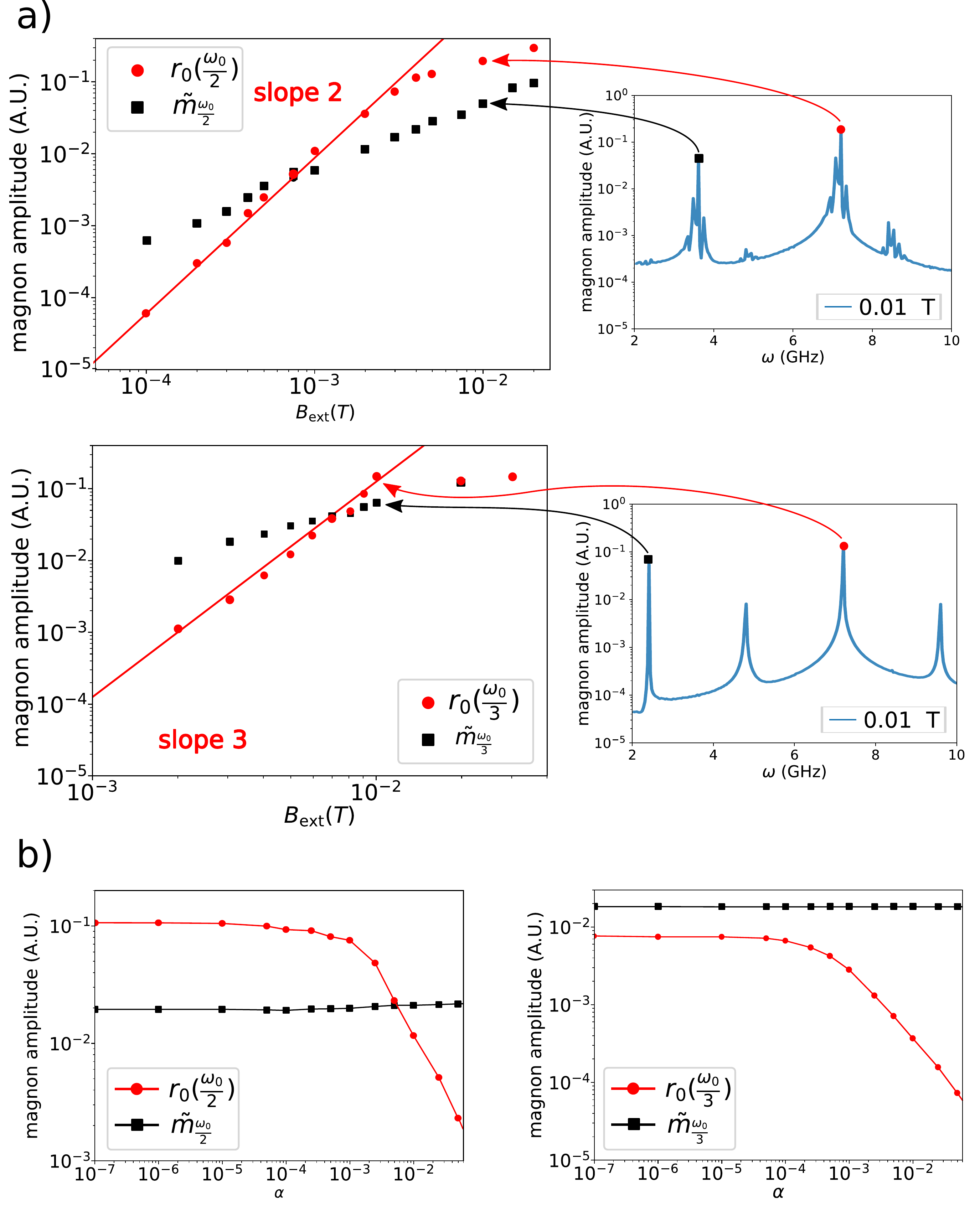}
    \caption{Plots of the amplitude of the excited modes in terms of a) the amplitude of the applied field with $\alpha = 10^{-3}$ and b) the damping parameter for an out-of-plane AC magnetic field $B_{\mathrm{ext}}=0.003$. In a) the straight lines with slope $2$ ($3$) indicate the growth with a power $2$ ($3$) for the second (third) harmonic generation. We notice that, above a certain amplitude of the applied field, the breathing mode is more excited than the mode with same frequency as the perturbation.  {The other two peaks on the inset of the third harmonic generation correspond to the second and forth harmonic generation. They do not resonate with any eigenmodes of the skyrmion.}}
    \label{fig:AmplitudeDependence}
    \end{center}
\end{figure}

We focussed on the excitation of the breathing mode by an out-of-plane field in order to analyze the amplitude of the eigenmode $r_{n}$, dependence on the applied field amplitude $B_{\textrm{ext}}$, and the material damping $\alpha$~\footnote{Notice that an out-of-plane AC field can only excite the breathing mode of an isolated radially symmetric skyrmion. Exciting the other modes requires a non-radially symmetric perturbation. This can either be done by other means or by out-of-plane fields accompanied by an additional radial symmetry breaking mechanism such as introducing temperature fluctuations.}. 
We applied out-of-plane AC magnetic fields with half and one third of the breathing mode frequency $\omega_0$. 
In Fig.~\ref{fig:AmplitudeDependence}a), as a function of the applied AC magnetic field strengths $B_{\mathrm{ext}}$ we show the amplitudes, extracted from the frequency spectrum (see right panel), of the breathing mode $r_0$ and the forced perturbation $\tilde{m}$ at the AC field frequency as a log-log plot.
The top (bottom) panel corresponds to $\omega_0/2$ ($\omega_0/3$) for a fix damping constant $\alpha= 10^{-3}$.  
While $\tilde{m}$ grows linearly with $B_{\mathrm{ext}}$, the amplitude of the breathing mode grows with a power $2$ for $\omega_{0}/2$ and with a power $3$ for $\omega_{0}/3$, in a certain range of the applied field, as expected for second and third harmonic generation. 
As the amplitude of the breathing mode grows beyond the linear approximation, the power law coefficient reduces due to further scattering of the magnons. 
Furthermore, as another important result of this manuscript, we find that above a certain amplitude of the applied field, the breathing mode is more excited than $\tilde{m}$.

In Fig.~\ref{fig:AmplitudeDependence}b), we show $r_0$ and $\tilde{m}$ as a function of the damping parameter $\alpha$ for $B_\mathrm{ext}= 0.003 \, \mathrm{T}$. Analogous to a), the left (right) panel corresponds to an applied AC magnetic field with frequency $\omega_0/2$ ($\omega_0/3$). 
We see that the forced perturbation amplitude $\tilde{m}$ is independent of the damping, as expected.
The breathing mode $r_0$, however, decreases as a function of $\alpha$. The Gilbert dissipation damps the energy flow from the forced perturbation to the resonant mode and reduces the amplitude from the resonant mode. Thus frequency multiplication is more efficient for systems with small damping.

\section{Discussion}
\label{sec:Discussion}

From the results above, we notice that the phenomenon of frequency multiplication might resemble the well-known parametric excitation~\cite{Zakharov1975,Bryant1988,Bertotti2001}. The main difference, however, is that the resultant frequency is a multiple and not a fraction of the applied frequency. This crucial difference presents considerable advantages of frequency multiplication for two main reasons: i) lower frequencies are easier to produce and guide experimentally, and ii) there are no magnons that are resonantly excited at the pumping frequency since the applied frequency is well below the magnon gap. The latter presents an experimental obstacle for parametric pumping since directly excited magnons at the pumping frequency are hard to avoid in a real experimental situation. These magnons can interact with the parametrically excited ones and, thus, hinder the ability to fully control the excited modes. Moreover, the excitation and scattering of magnons with higher frequency during parametric excitation are very sensitive to sample inhomogeneities.

To summarize, we have proposed the excitation of the eigenmodes of topological magnetic textures by frequency multiplication based on the general non-linear properties of magnetization dynamics. While we have explicitly demonstrated this effect by means of micromagnetic simulations on the excitation modes of isolated magnetic skyrmions, our theoretical analysis reveals its universal behavior, i.e.\ being independent of microscopic details of the magnetic structures as well as the source of the perturbation.
Not only does this provide a new method to excite the eigenmodes by applying lower frequency amplitudes, this frequency multiplication mechanism propounds novel applications in spintronics devices such as an in-material frequency multiplier for magnonic applications~\cite{Dussaux2010,Finocchio2015,Wagner2016,Hamalainen2018,Hollander2018}. In particular, bound modes not excited via linear resonance and which are weakly coupled (e.g. by dipolar stray fields or spin-wave excitations) serve as a building block for so-called parametrons, a computing scheme which has seen a strong revival in the recent discussions of oscillator-based computing~\cite{Csaba2020}. Furthermore, the tunability of the magnetic textures by altering the material properties, temperature and applied fields make them very versatile.

\section{Acknowledgement}
\label{sec:Acknowledgement}

We thank V.K. Bharadwaj, Ar. Abanov, O. Gomonay and T. Br\"acher for discussions.
We acknowledge funding from the Deutsche Forschungsgemeinschaft (DFG, German Research Foundation): TRR 173 – projects 268565370 (B01) and 268565370 (B12),
SPP 2137 – projects 403233384 and 403512431 as well as the Emmy Noether Project of K.E.S., project number 320163632. 
D.R. and K.E.S acknowledge funding from the Emergent AI Center funded by the Carl-Zeiss-Stiftung and from JGU TopDyn. 
J.N. acknowledges support from the Max Plack Graduate Center. 
This research was supported in part by the National Science Foundation under Grant No. NSF PHY-1748958.

\end{document}